\documentstyle[prl,aps,twocolumn,epsf]{revtex}

\draft

\begin{document}

Comment on "Level Statistics of Quantum Dots Coupled to Reservoirs"

In a recent letter, K\"onig {\it et al.} describe how two levels of a quantum dot are broadened by the coupling to an 
external reservoir\cite{Konig}. To do this they consider  that these two levels of the dot are nearest neighbors in energy and are well separated 
from other levels. They find that when the coupling to the reservoir increases, the spectral weight of one of these two states 
$\sigma$, after being roughly a Lorentzian when the width $\Gamma_\sigma$ is smaller than the distance $\Delta\epsilon$ between the two 
levels,  evolves into a structure made of a sharp peak located at the  center of the two levels plus a 
broad contribution. Since each  peak moves towards the middle of the original levels, this motion is analyzed in term of level attraction.
 We show here that the conclusions of ref.[1] are very specific to the case of two levels. Considering 
more than two levels lead to a different description, namely each spectral weight is actually splitted into {\it several peaks} so that it is meaningless analyze the results in terms of level attraction. We consider the same hamiltonian as in ref.\cite{Konig}
\begin{equation} H =  \sum_k\epsilon_k a^+_k a_k + \sum_{\sigma=1,m}\epsilon_\sigma a^+_\sigma a_\sigma 
+ \sum_{k \sigma} [T_{k \sigma} a^+_k c_\sigma + h.c.]
\end{equation}
This generalizes the hamiltonian considered in ref.\cite{Konig} to $m$ discrete levels coupled to a continuum.
The spectral weight of a given level $\sigma$ is given by 
$A_\sigma(\omega)= - \mbox{Im} G_{\sigma\sigma}(\omega+i 0^+)/\pi$. The Green's functions $G_{\mu\sigma}$ obey the $m \times m$ 
matrix equation:
$\sum_\mu M_{\lambda \mu} G_{\mu \sigma} = \delta_{\lambda,\sigma}
$
where $M_{\lambda \mu}=(\omega - \epsilon_\lambda)\delta_{\lambda\mu}-\Sigma_{\lambda\mu}$, with
$\Sigma_{\lambda\mu}= -i \pi \sum_k T_{k\lambda} T^*_{k\mu} \delta(\omega - \epsilon_k)$. Since the density of states in 
the reservoirs forms a continuum, $\Sigma_{\lambda\mu}$ can be rewritten as $\Sigma_{\lambda\mu}=- i \pi \langle 
T_{k\lambda} 
T^*_{k\mu} \rangle \rho_0$, where  the density of states of the continuum, $\rho_0$, is  supposed to be constant. In the case considered in ref.\cite{Konig} where the tunnel matrix elements are assumed to be 
independent of the reservoir state, $T_{k\sigma}=T_\sigma$, the above matrix can be diagonalized and the Green's function 
can be calculated. We find:
\begin{equation}
G_{\sigma\sigma}(\omega)=g_{\sigma}(\omega){2 + i \sum_{\lambda\neq\sigma}\Gamma_\lambda \ g_{\lambda}(\omega)
\over 2 + i \sum_{\lambda}\Gamma_\lambda \ g_{\lambda}(\omega)}
\label{green}
\end{equation}
where $\Gamma_{\lambda}= 2 \pi |T_\lambda|^2 \rho_0$ is the width of a level without coupling between levels. $g_\lambda(\omega)=(\omega-\epsilon_\lambda)^{-1}$. The complex eigenvalues 
$\{ \tilde\epsilon_\alpha + i \tilde\Gamma_\alpha\}$ are solutions of \begin{equation}\sum _\lambda{
 \Gamma_\lambda /2\over
\tilde\epsilon_\alpha + i \tilde\Gamma_\alpha- \epsilon_{\lambda}
}
= i
\label{solutions}
\end{equation}

When the $\Gamma_\lambda$ increase and become large compared to the average distance between levels $\Delta$, the spectral weight $A_{\sigma}(\omega)$ of each level is splitted into {\it a series} of $(m-1)$ peaks  whose centers are given by:
$
\sum _\lambda
(\tilde\epsilon_\alpha - \epsilon_{\lambda} )^{-1}
= 
0
$
and are placed at intermediate positions between the unperturbed levels (Fig. 1.a,b).
Their widths $\tilde\Gamma_\alpha$ are given of order of $( \sum_\lambda (\Gamma_\lambda/2)/(\epsilon_{\lambda}-\tilde\epsilon_\alpha)^2)^{-1} \sim \Delta^2 / 
\Gamma$ where $\Gamma$ is the typical value of a bare level.
The corresponding spectral weights are $A^\alpha_{\sigma}=
1 / [(\tilde\epsilon_\alpha-\epsilon_{\sigma})^2 
\sum_\lambda 1 / (\epsilon_{\lambda} - \tilde\epsilon_\alpha)^2] 
$
The $(m-1)$ peaks actually carry  $(1-1/m)$ of the spectral weight.
The  rest is carried by a very broad Lorentzian of width $\sum_\lambda \Gamma_\lambda$.  When two neighboring levels are well separated from others the spectral weight tends to concentrate on only one peak, the situation considered in ref.\cite{Konig}. However as it is shown on  Fig. 1c, the weight of other peaks may be not negligible.\vspace{-0.5cm}
\begin{figure}[h]
\centerline{
\epsfxsize 8cm
\epsffile{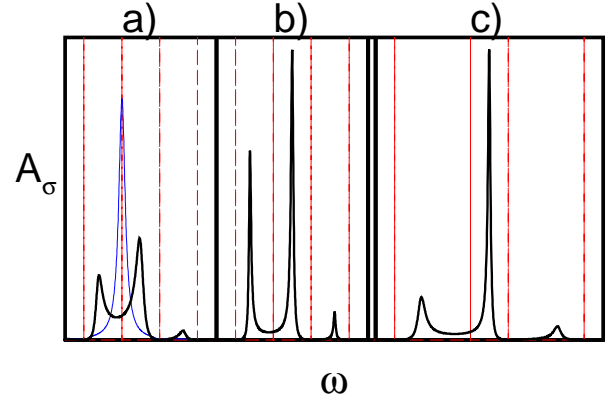}}
\caption{\protect\small Spectral weight $A_\sigma(\omega)$ for the $2^{nd}$ out of $4$ discrete levels whose unperturbed energies are figured by vertical lines. All the widths $\Gamma_\sigma$ are taken equal $\Gamma_\sigma=\Gamma$. a) $\Gamma / \delta=0.1$ (thin lines) $\Gamma / \delta=1$. b) $\Gamma / \delta=3$. c) $\Gamma / \delta=3$, here the bare levels are not equidistant ($2\delta, \delta, 2\delta$). $\delta$ is  the distance between the $2^{nd}$ and the $3^{rd}$ levels.
}
\label{spectralweight}
\end{figure}

\medskip
Finally we note that this structure is extremely sensitive to the type of coupling to the reservoirs. 
Refs.\cite{Konig,Pascaud} assume a constant coupling $T_{k\sigma}=T_\sigma$. The most general case
$\langle T_{\lambda_i k}T_{k \lambda_j} \rangle \neq
\langle T_{\lambda_i k} \rangle \langle T_{k \lambda_j} \rangle$, 
has also been discussed in ref\cite{Pascaud}.  When the  couplings to the reservoir levels are uncorrelated and symmetric so 
that 
$\langle T_{\lambda_i k}T_{k \lambda_j} \rangle =
\langle T_{\lambda_i k} \rangle \langle T_{k \lambda_j} \rangle=0 $ 
for $i \neq j$, 
the off-diagonal elements in the matrix $M$ vanish and one finds that
 each level coupled to the continuum
is broadened into a Lorentzian according to the Fermi golden rule; neither its center nor
its width are altered:
``the resonances do not talk to each other''. The coupling between resonances is thus critically dependent on the type of coupling with the reservoirs.

\medskip

M. Pascaud and G.
Montambaux

Laboratoire de Physique des Solides,  associ\'e au
CNRS, 

Universit\'{e} Paris--Sud, 91405 Orsay, France.

\end{document}